\documentclass[12pt,preprint]{aastex}
\usepackage[dvips]{color}
\usepackage{graphicx}

\def\msol{\hbox{\kern 0.20em $M_\odot$}}
\def\lsol{\hbox{\kern 0.20em $L_\odot$}}
\def\rsol{\hbox{\kern 0.20em $R_\odot$}}
\def\sr{\hbox{\kern 0.20em sr}}
\def\srmu{\hbox{\kern 0.20em sr$^{-1}$}}
 
\def\g{\hbox{\kern 0.20em g}}
\def\gmu{\hbox{\kern 0.20em g$^{-1}$}}
\def\kg{\hbox{\kern 0.20em kg}}
\def\pc{\hbox{\kern 0.20em pc}}
 
\def\mum{\hbox{\kern 0.20em $\mu$m}}
\def\mumd{\hbox{\kern 0.20em $\mu$m$^{-2}$}}
\def\cm{\hbox{\kern 0.20em cm}}
\def\m{\hbox{\kern 0.20em m}}
\def\km{\hbox{\kern 0.20em km}}
\def\nm{\hbox{\kern 0.20em nm}}
 
\def\s{\hbox{\kern 0.20em s}}
\def\h{\hbox{\kern 0.20em h}}
\def\sec{\hbox{\kern 0.20em sec}}
\def\min{\hbox {\kern 0.20em min}}
\def\smu{\hbox{\kern 0.20em s$^{-1}$}}
\def\smd{\hbox{\kern 0.20em s$^{-2}$}}
\def\an{\hbox{\kern 0.20em an}}
\def\anmu{\hbox{\kern 0.20em an$^{-1}$}}
\def\deg{\hbox{\kern 0.20em $^{\rm o}$}}
\def\yr{\hbox{\kern 0.20em yr}}
\def\yrmu{\hbox{\kern 0.20em yr$^{-1}$}}
\def\Myr{\hbox{\kern 0.20em Myr}}
\def\Mymu{\hbox{\kern 0.20em Myr$^{-1}$}}
\def\K{\hbox{\kern 0.20em K}}
\def\pcmu{\hbox{\kern 0.20em pc$^{-1}$}}
\def\pcmd{\hbox{\kern 0.20em pc$^{-2}$}}
\def\pcmt{\hbox{\kern 0.20em pc$^{-3}$}}
\def\kms{\hbox{\kern 0.20em km\kern 0.20em s$^{-1}$}}
\def\kmpd{\hbox{\kern 0.20em km$^{2}$}}
\def\kpc{\hbox{\kern 0.20em kpc}}
\def\cms{\hbox{\kern 0.20em cm\kern 0.20em s$^{-1}$}}
\def\erg{\hbox{\kern 0.20em erg}}
\def\ergs{\hbox{\kern 0.20em erg}}
\def\cmpd{\hbox{\kern 0.20em cm$^2$}}
\def\cmmd{\hbox{\kern 0.20em cm$^{-2}$}}
\def\cmms{\hbox{\kern 0.20em cm$^{-6}$}}
\def\cmpt{\hbox{\kern 0.20em cm$^3$}}
\def\cmmt{\hbox{\kern 0.20em cm$^{-3}$}}
\def\mpd{\hbox{\kern 0.20em m$^2$}}
\def\mmd{\hbox{\kern 0.20em m$^{-2}$}}
\def\mpt{\hbox{\kern 0.20em m$^3$}}
\def\mmt{\hbox{\kern 0.20em m$^{-3}$}}
\def\mujy{\hbox{\kern 0.20em $\mu$Jy}}
\def\mjy{\hbox{\kern 0.20em mJy}}
\def\Mj{\hbox{\kern 0.20em MJy}}
\def\jy{\hbox{\kern 0.20em Jy}}
\def\ghz{\hbox{\kern 0.20em GHz}}
\def\srmd{\hbox{\kern 0.20em sr$^{-1}$}}
\def \kms{km~$\rm{s}^{-1}$}

\def \mum{$\mu$m}

\def\G{\hbox{\kern 0.20em G}}

\def\h13cop{\hbox{H$^{13}$CO$^{+}$}}

\def\S+{\hbox{S{\small II}}}

\slugcomment{The Evolving ISM in the Milky Way \& Nearby Galaxies}

\shorttitle{ISMEVOL}
\shortauthors{Noriega-Crespo et al.}

\begin{document}

\newcommand{\jfourteen}{\hbox{$J=14\rightarrow 13$}}
 \title{Spitzer Observations of Supernova Remnant IC443}

\author{A.\ Noriega-Crespo\altaffilmark{1},
D. C. Hines\altaffilmark{2},
K. Gordon\altaffilmark{3,4}, 
F. R. Marleau\altaffilmark{1}, 
G. H. Rieke\altaffilmark{3}, 
J. Rho\altaffilmark{1},
\& W. B. Latter\altaffilmark{5}}

\altaffiltext{1}{SPITZER Science Center, California Institute of
Technology, Pasadena, CA 91125}
\altaffiltext{2}{Space Science Institute, Corrales, NM, 87048}
\altaffiltext{3}{Steward Observatory, University of Arizona, Tucson, AZ, 85721}
\altaffiltext{4}{Space Telescope Institute, Baltimore, MD, 21218}
\altaffiltext{5}{NASA Herschel Science Center, Pasadena, CA, 91125}

\begin{abstract}
We present Spitzer observations of IC~443 obtained with MIPS and IRS
as part of our GTO program on the astrophysics of ejecta from evolved stars.
We find that the overall morphology at mid/far IR wavelengths resembles even 
more closely a loop or a shell than the ground based optical and/or near
IR images.The dust temperature map, based on the 70/160\mum~ratio, shows a 
range from 18 to 30 K degrees. The IRS spectra confirm the findings from 
previous near+mid IR spectroscopic observations of a collisionally excited 
gas, atomic and molecular, rich in fine structure atomic and pure H$_2$ 
rotational emission lines, respectively. The spectroscopic shock indicator, 
[Ne~II] 12.8\mum, suggests shock velocities ranging from 60-90 \kms, 
consistent with the values derived from other indicators.
\end{abstract}

\keywords{infrared: ISM --- ISM: individual (IC 443) --- supernova remnants}


\section{Introduction}
As one of the best examples of a supernova remnant (SNR) interacting with a 
molecular cloud IC~443 has been studied over all possible wavelength ranges, 
from the radio (see e.g. Leahy 2004), through the sub-mm (van Dishoeck et al. 
1993) to the X-rays (see e.g. Troja et al. 2006, 2008), including  
TEV $\gamma$ emission that is thought to be associated with pulsars 
(Albert et al. 2007; Humesky et al. 2007). 
At an estimated distance of 1.5 Kpc (Welsh \& Sallmen 2003), 
IC~443 covers approximately a square degree over the sky. Until recently 
because of its relatively large size, most of IC~443 imaging data was a 
by-product of large sky surveys (IRAS, 2MASS, ROSAT, MSX, etc), and to this 
date the spectroscopic data only samples a handful of specific regions.
The spectroscopic data, nevertheless, do confirm that the emission arising from
IC~443 carries the signature of collisionally excited (atomic \& molecular) 
gas, the result of a shock wave impinging on a nearby molecular cloud.
(see e.g. Shull et al 1982; Graham et al. 1987, Burton 1987, van Dishoeck et 
al. 1993; Cesarsky et al. 1999, Oliva et al. 1999; Rho et al. 2001, Neufeld 
et al. 2007, Rosado  et al. 2007, among others). 
Thus IC~443 continues to provide a excellent laboratory to study the evolution
and interaction of a SNR with its surrounding medium.

 In this communication we present the images obtained with the far infrared 
(FIR) photometer MIPS (Rieke et al. 2004), complemented with mid infrared (MIR)
spectroscopy data obtained with IRS (Houck et al. 2004), both instruments on 
board of the Spitzer Space Telescope (Werner et al. 2004).

\begin{figure}
\centerline{
\includegraphics[width=562pt,height=401pt,angle=0]{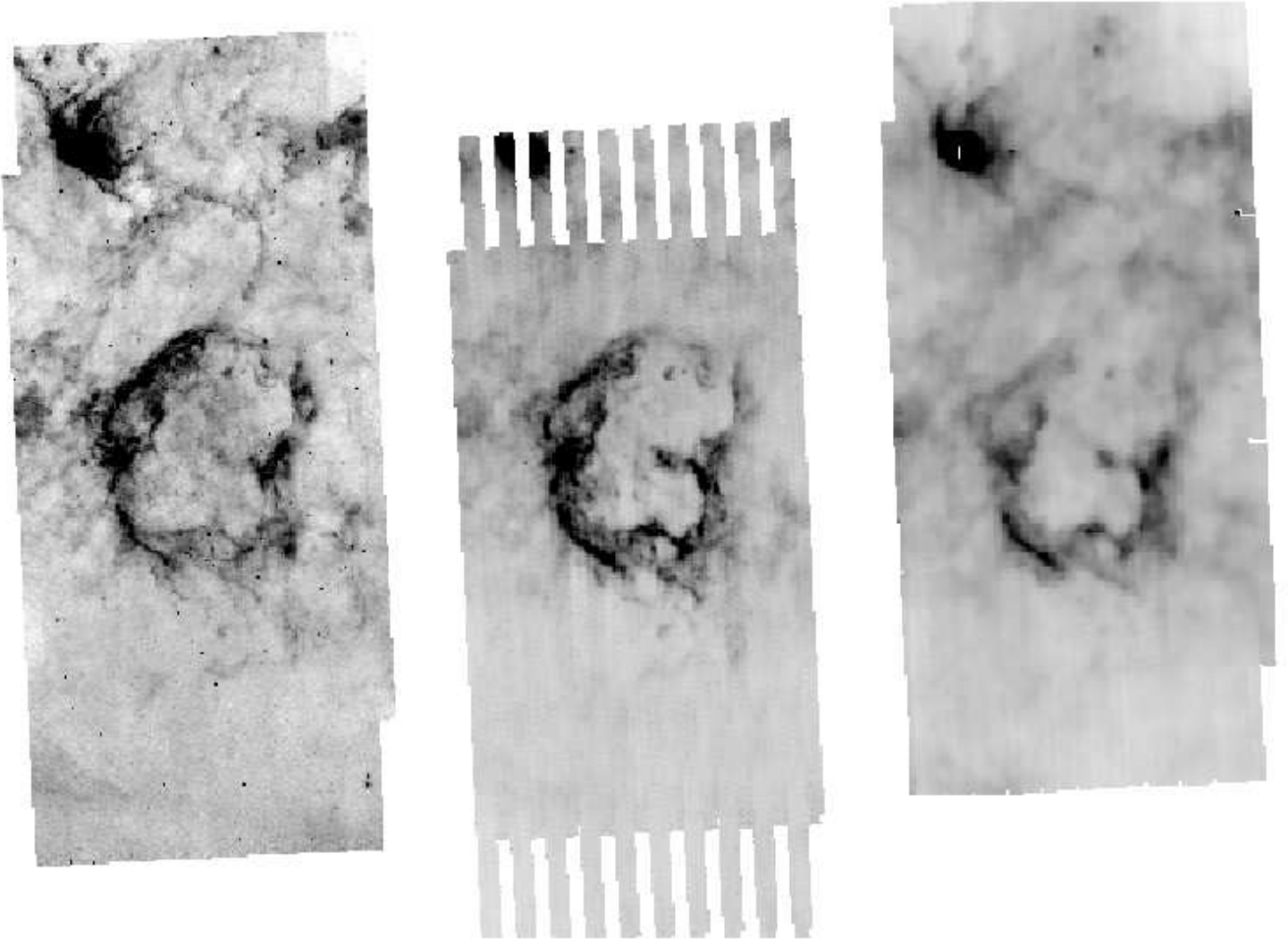}}
\caption{\label{fig:1} From left to right: MIPS maps of IC 443 at 
24, 70 \& 160\mum. FOV$\sim$0.9\arcdeg$\times$1.9\arcdeg. The bright source at
the top of the image is IC~444 or IRAS 0655+2319. North is up and East is left}
\end{figure}

\section{Observations} 

The MIPS \& IRS observations are part of our GTO program (Rieke PID 77 and 
Houck PID 18) to study the physical characteristics of the ejecta from evolved
stars. Although the observations were taken very early in the Spitzer mission,
we have learned a handful of new things on data reduction as 
to provide the best possible images and spectra.
The MIPS observations were obtained at three different epochs using fast
scan mapping (3 sec per frame, 5 pointings per pixel), with scan legs offset
of 148" to sample completely the 70 and 160um arrays. One of the remarkable 
features of the MIPS instrument is its capability to map large areas of the 
sky in a very efficient way, and therefore the new MIPS images at 24, 70 \& 
160\mum~capture the SNR in its entirety (Fig. 1). The IRS observations were 
carried out using both short and long high resolution modules at 5 fixed 
cluster positions (including an off-position) using 6 and 14sec ramps (one 
cycle), respectively. The off-position was used to remove the background from 
the on-target spectra.

\begin{figure}

\centerline{
\includegraphics[width=560pt,height=260pt,angle=0]{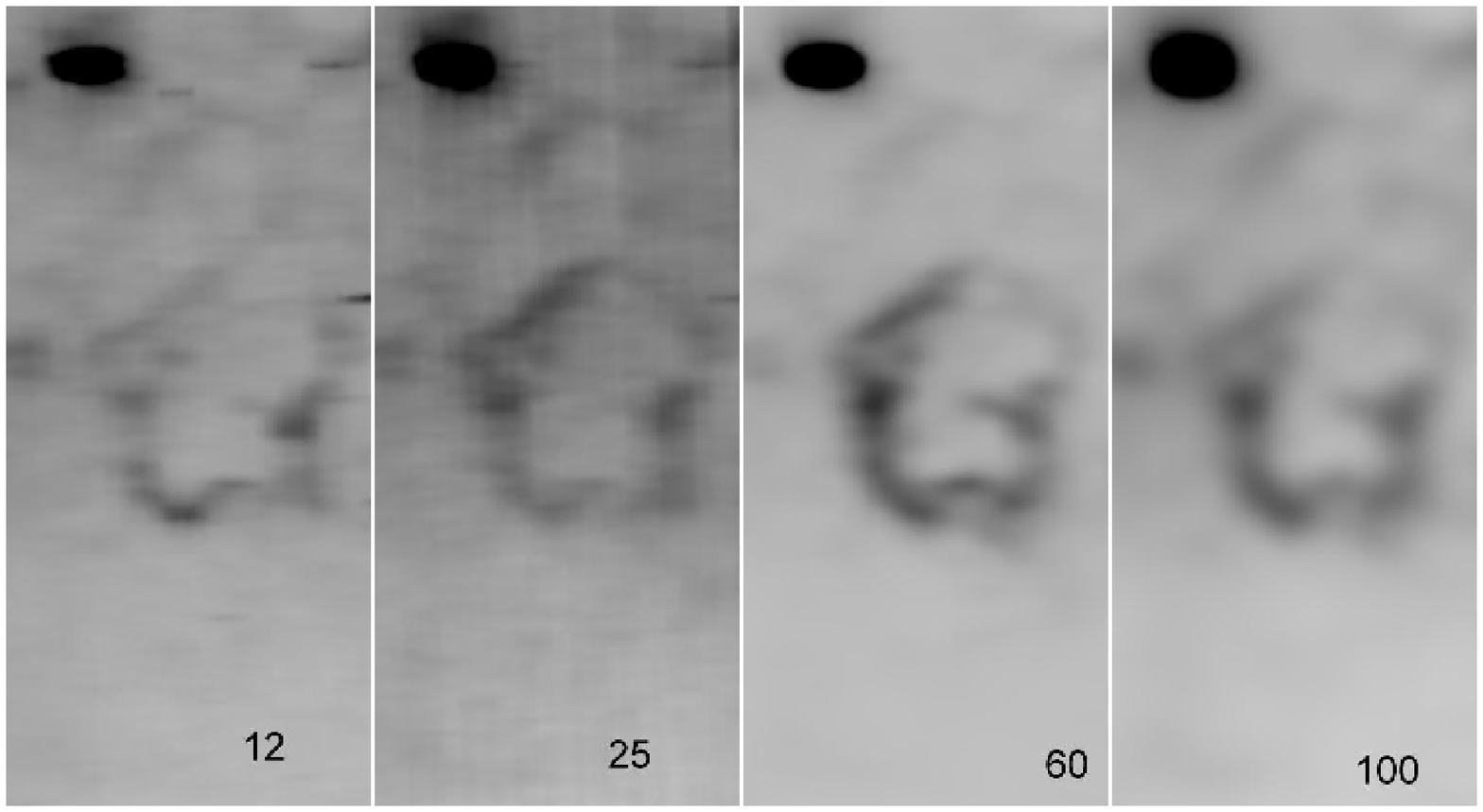}}
\centerline{
\includegraphics[width=305pt,height=270pt,angle=0]{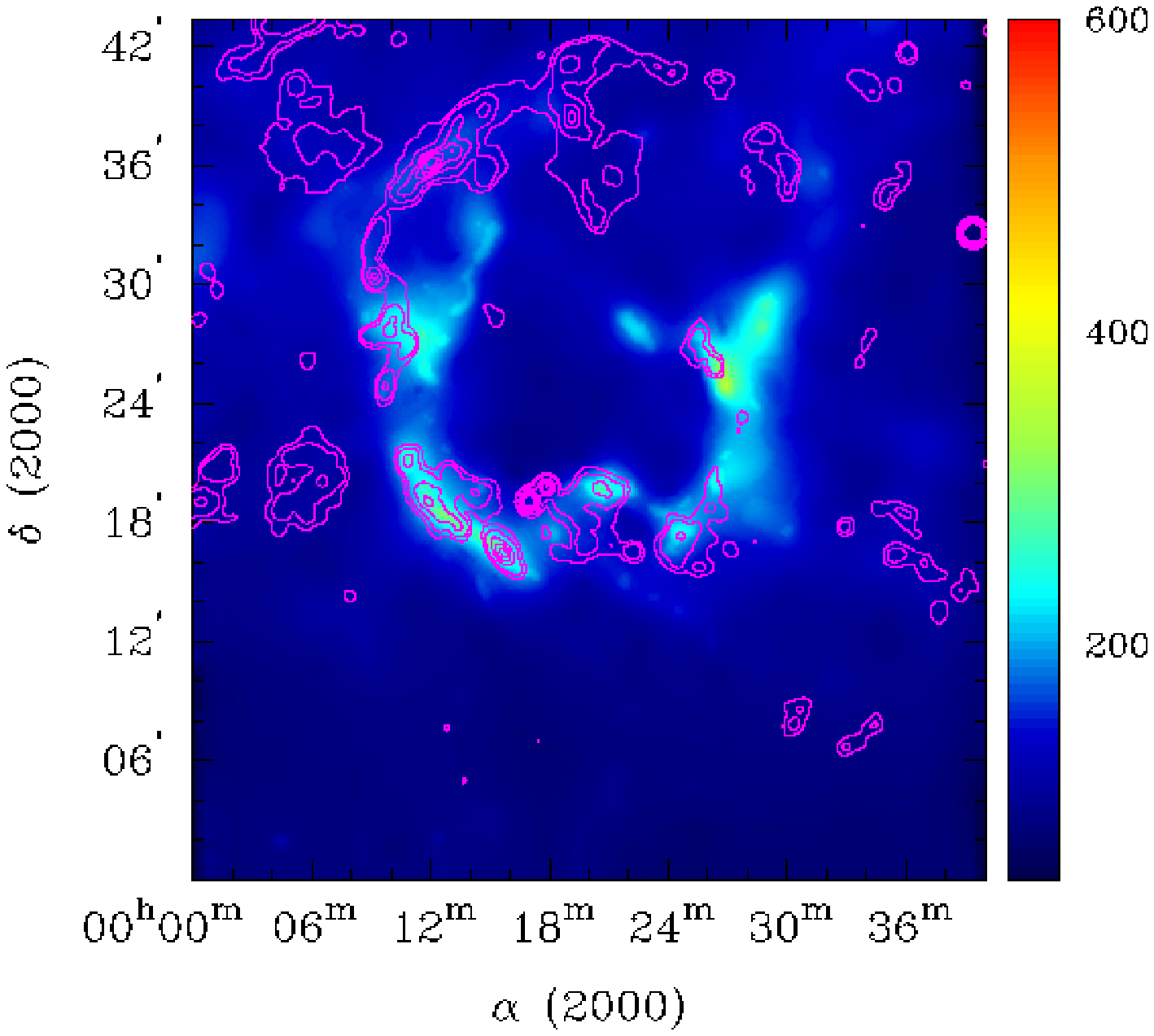}
\includegraphics[width=305pt,height=270pt,angle=0]{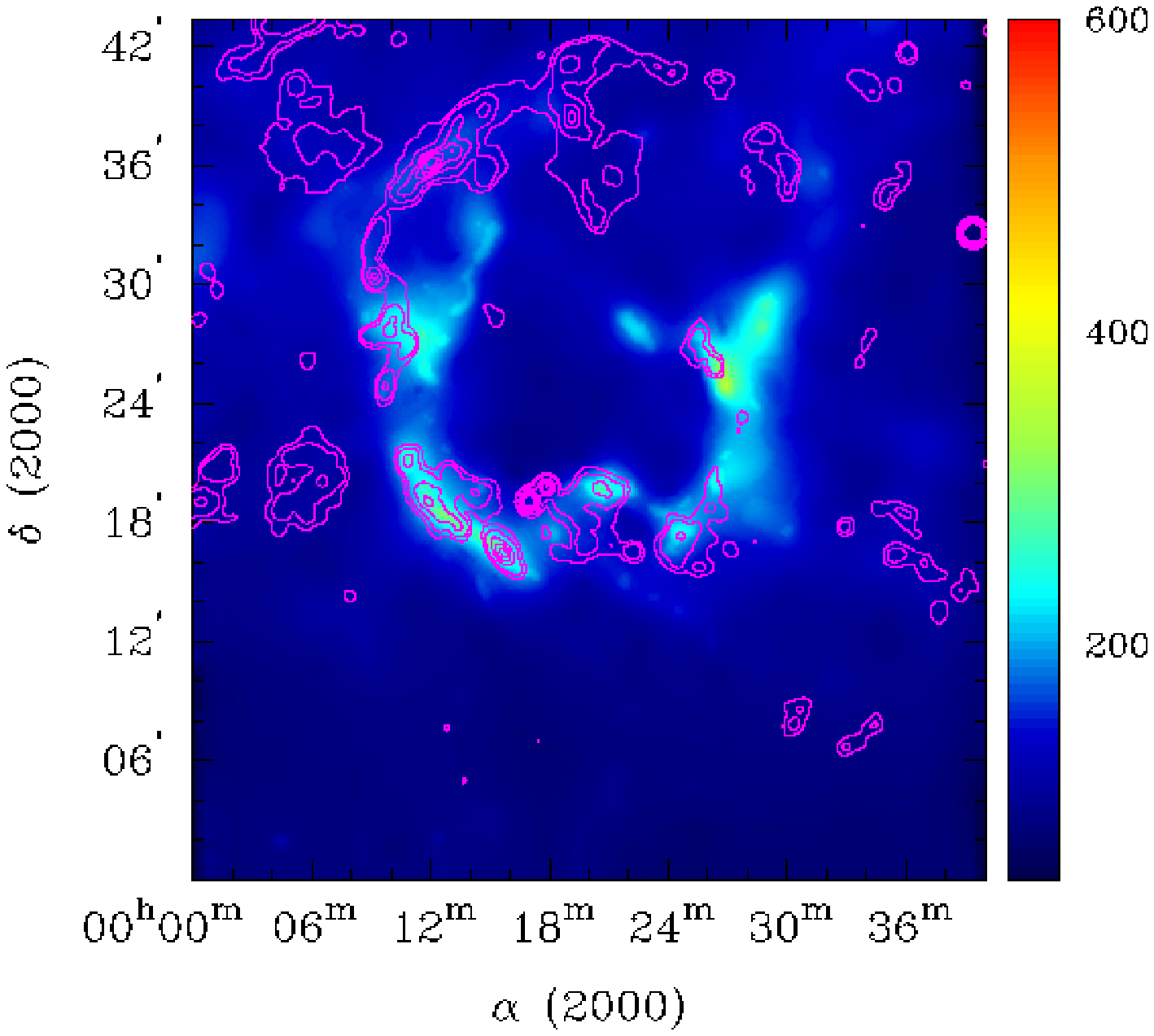}}
\caption{\label{fig:2} Top; IRAS HiRes fresco of IC~443 with a similar
FOV as Fig 1. Bottom Left: H$\alpha$ (false color) and MIPS 24\mum~(contours).
Rigft: MIPS 160\mum~(false color) and HI 1.4 GHz (contours). The color scales 
are in MJy/sr and the FOV $\sim$ 0.9\arcdeg~radius}
\end{figure}

\begin{figure}
\centerline{
\includegraphics[width=280pt,height=280pt,angle=0]{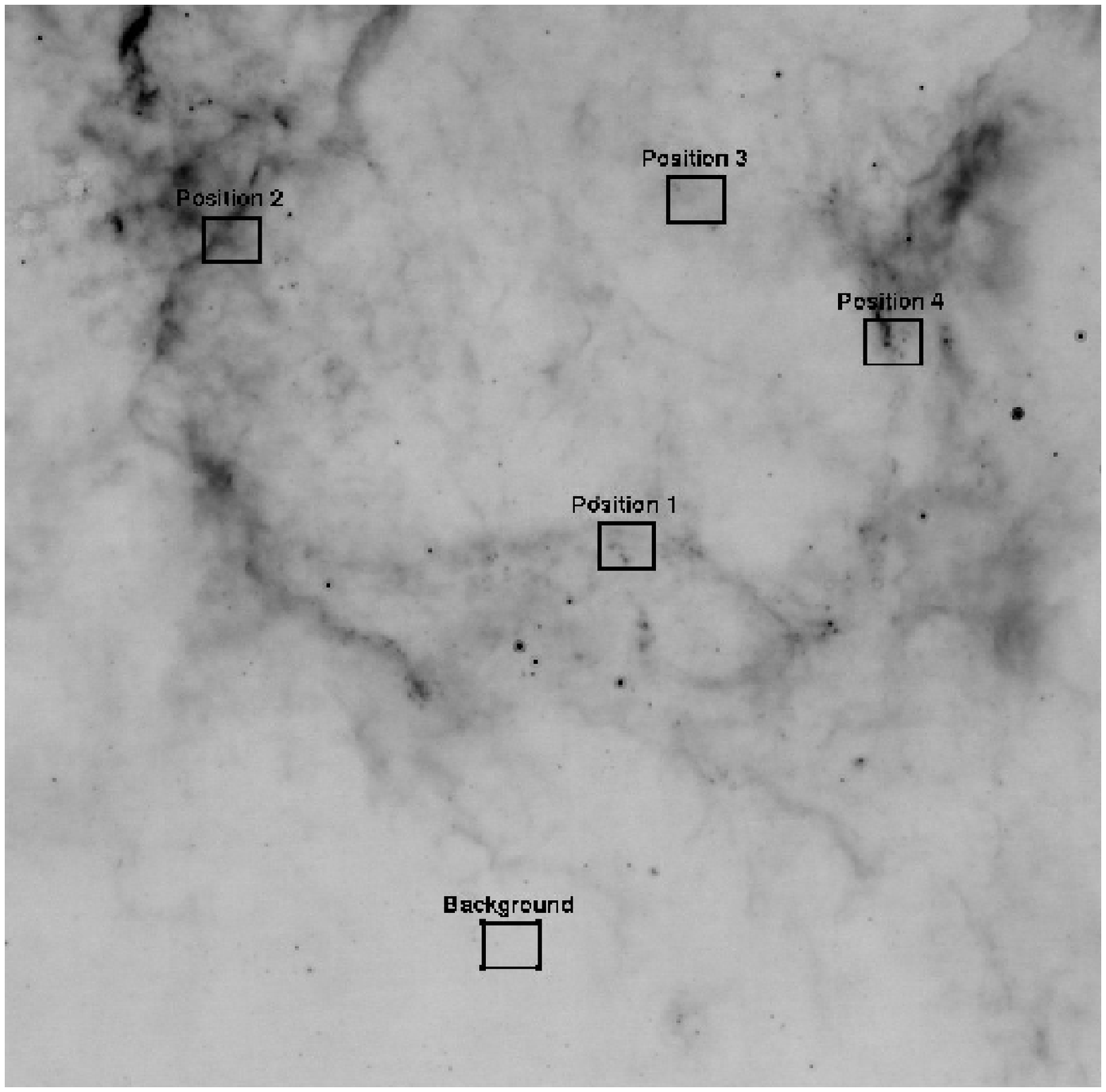}}
\centerline{
\includegraphics[width=280pt,height=280pt,angle=90]{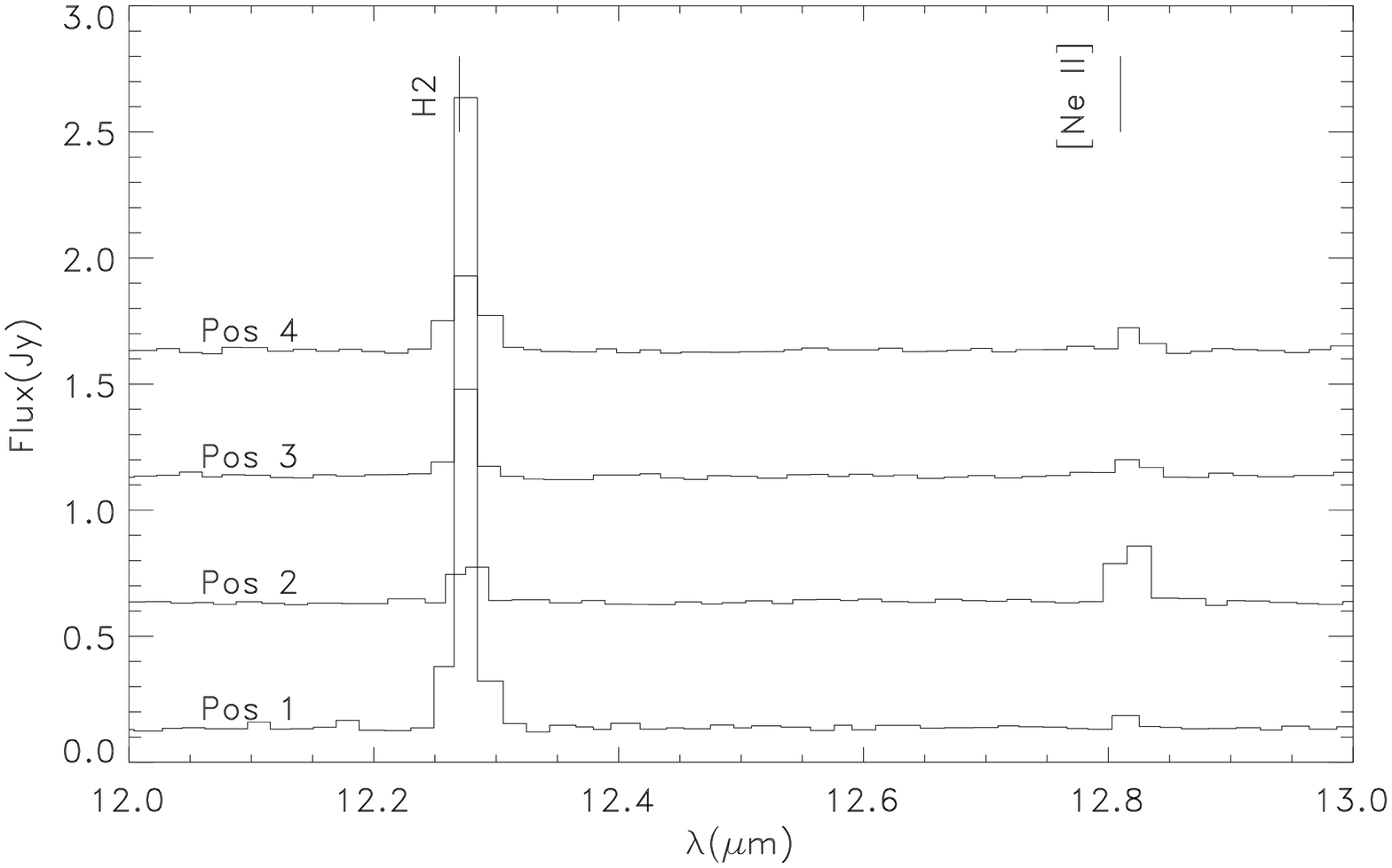}
\includegraphics[width=280pt,height=280pt,angle=90]{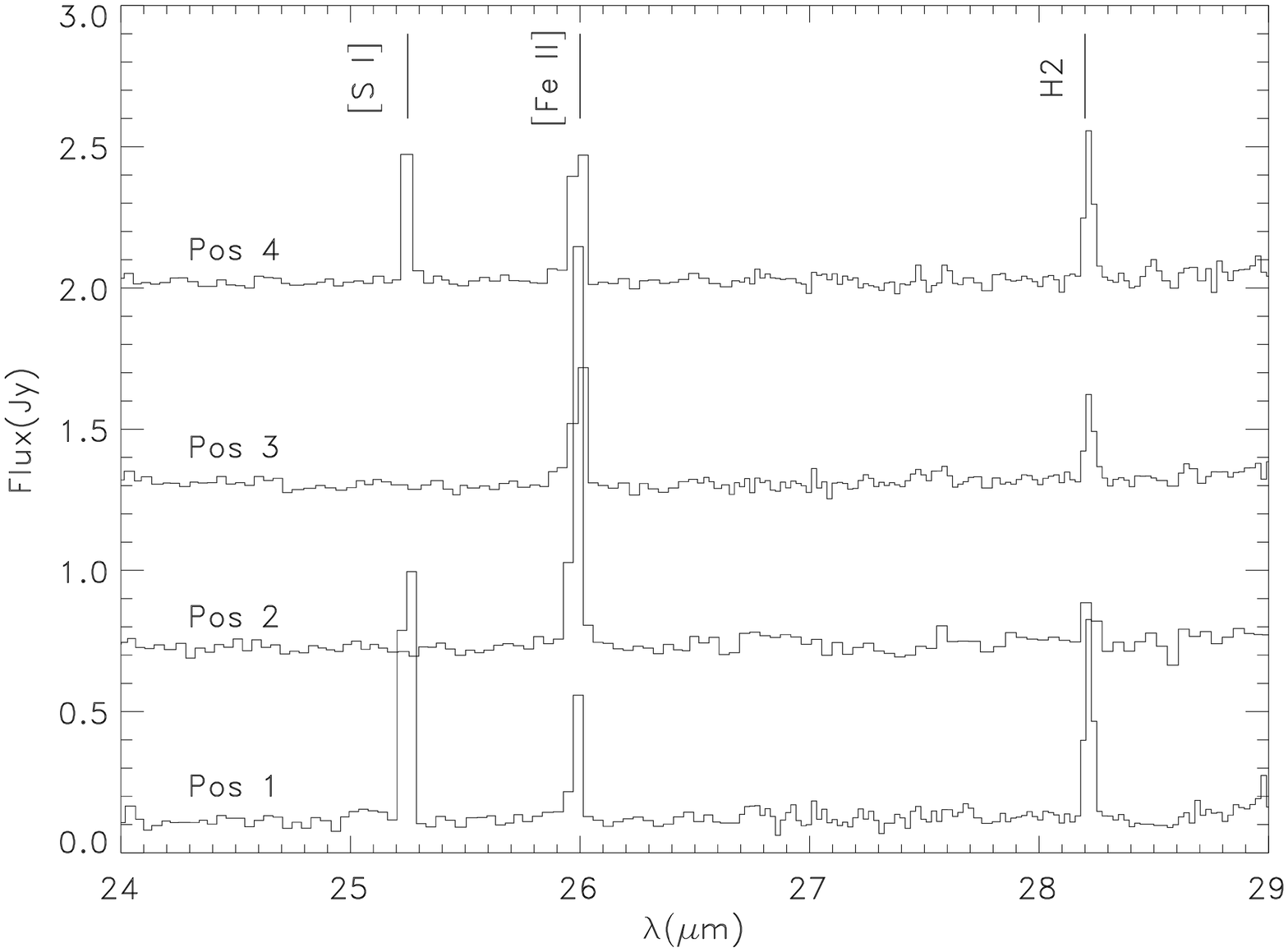}}
\caption{\label{fig:3} Top: A schematic view of the 5 IRS observed positions.
Bottom: Sample spectra obtained with the IRS short \& high resolution modules.}
\end{figure}

\section{Preliminary Analysis and Summary}

The morphology of the IC~443 is shown in superb detail in the 
high angular resolution MIPS images (standard beam sizes of 6\arcsec, 
18\arcsec~and 40\arcsec~at 24, 70 \& 160\mum~respectively). 
Nevertheless the overall shell 
morphology can be seen already in the IRAS images (Fig. 2 top, HiRes fresco 
first iteration; see also Braun \& Strom 1986). The comparison with
H$\alpha$ (a tracer of the ionized gas) and 24\mum~confirms that a significant
fraction of the emission at 24\mum~ is due to fine structure atomic and 
H$_2$ molecular emission lines, and not necessarily to dust continuum emission 
from small dust grains. his conclusion is further supported by the IRS 
spectra (Fig. 3), which show strong [Fe~II] 26\mum~and 
H$_2$ 0-0 S(0) 28.2\mum~emission lines at the four observed positions, 
but no detected continuum emission.Indeed, except for the South Rim of the 
shell, the 160\mum~emission (a tracer of cold dust) does not match the 
morphology of the HI 1.4GHz emission (Fig. 2, bottom left), suggesting that 
a large fraction of the emission is not due to dust continuum. The 2MASS Ks 
observations at 2\mum~were interpreted as due to H$_2$ excitation from shocks 
(Rho et al. 2001), if this is the case, then is possible that [C~II] 158\mum~
contribute to the 160\mum~emission. Certainly [O~I] 63\mum~has been detected 
in several positions across the shell (Rho et al. 2001), and is very likely to 
contribute significantly to the 70\mum~emission band. Even so, one can use the
70 to 160\mum~ratio to estimate the dust temperature, and at first 
approximation, we found a range of 18$-$30 K, with higher dust temperature 
at the NE, where the H$\alpha$ and 24\mum~emission are brighter.

\begin{figure}[t]
\centerline{
\includegraphics[width=250pt,height=180pt,angle=0]{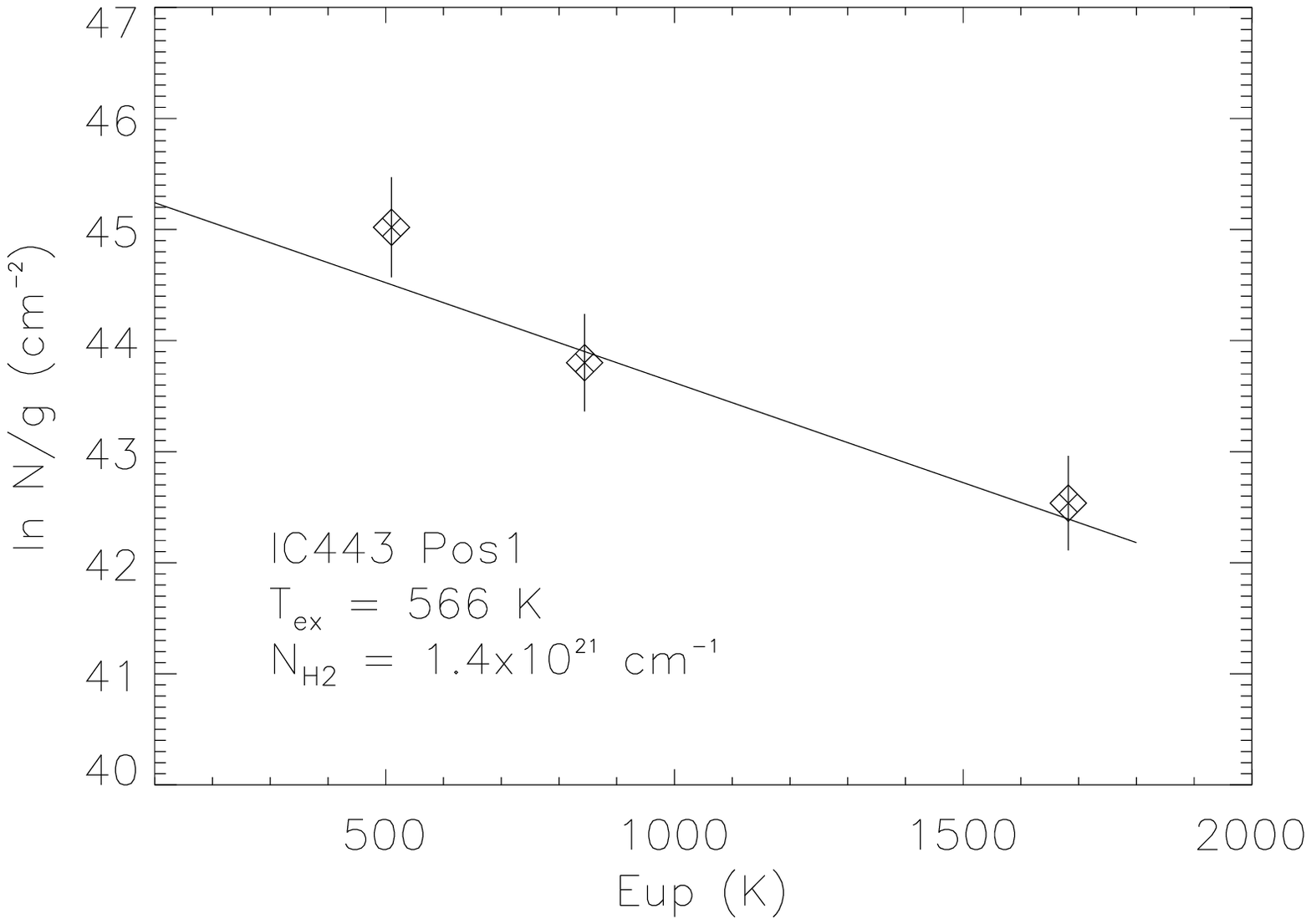}
\includegraphics[width=250pt,height=180pt,angle=0]{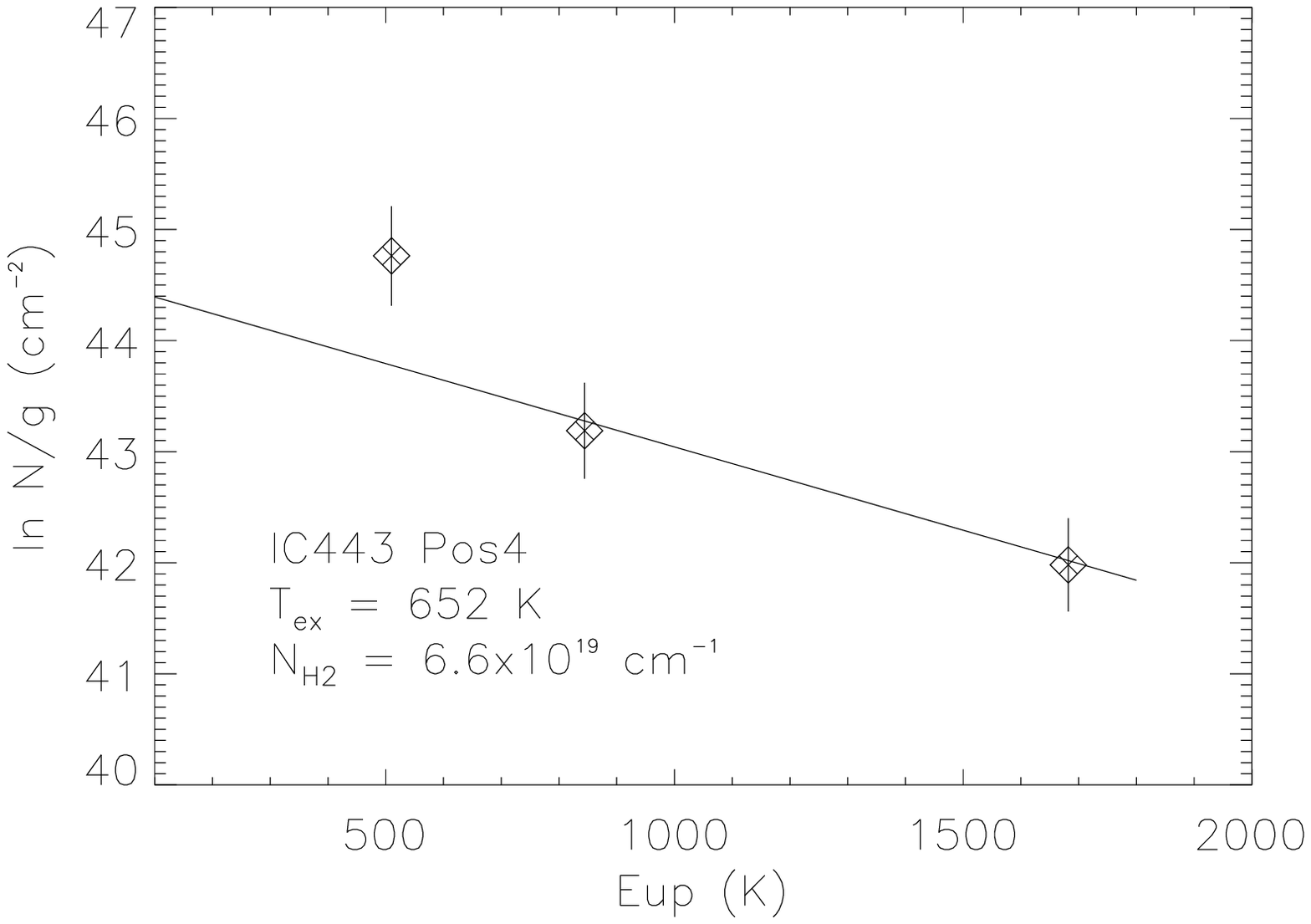}}
\caption{\label{fig:4} Excitation diagrams at positions one (left)
and four (right) based on the three H$_2$ 0-0 lines
(12.23, 17.03 and 28.22\mum) present within the wavelength range
of our IRS spectra.}
\end{figure}

The IRS spectra, as expected from previous work in the NIR+MIR, contains 
a handful of atomic fine structure lines from Fe, Ne and Si, plus the H2 
pure rotational lines (Fig. 3, bottom). The most interesting aspect is
the obvious differences as a function position in the excitation along the 
shell. The standard shock indicator of [Ne~II] 12.8\mum~suggests shock 
velocities ranging from 60-90 km/s, and consistent with some previous 
estimates to account for the emission of the atomic/ionic lines 
(Rho et al.2001).

Finally, the excitation diagrams derived from the three H2 lines covered 
by the IRS observations (12.23, 17.03 and 28.22\mum) do also show differences 
in column densities and temperatures as a function of position, ranging
from T$_{ex}\sim 300 - 600$ K and 
N$_{H2}\sim 6.6\times 10^{19} - 1.4\times 10^{21}$ cm$^{-1}$ (Fig. 4)
suggesting that the interaction between the shock wave and its environment
is non-symmetric.



\end{document}